\documentstyle[aasms4]{article}
\makeatletter
  
  \@addtoreset{equation}{section}
\makeatother

\begin{document}

\title{RELATIVISTIC CORRECTIONS TO THE SUNYAEV-ZELDOVICH EFFECT FOR CLUSTERS OF GALAXIES. V. PRECISION ANALYTIC FITTING FORMULA FOR THE CROSSOVER FREQUENCY REGION}

\author{NAOKI ITOH AND YOUHEI KAWANA}

\affil{Department of Physics, Sophia University, 7-1 Kioi-cho, Chiyoda-ku, Tokyo, \\
102-8554, Japan; n\_itoh, y-kawana@hoffman.cc.sophia.ac.jp}

\centerline{AND}

\author{SATOSHI NOZAWA}

\affil{Josai Junior College for Women, 1-1 Keyakidai, Sakado-shi, Saitama, \\
350-0295, Japan; snozawa@josai.ac.jp}

\begin{abstract}

  We have succeeded in obtaining a precision analytic fitting formula for the exact numerical results of the relativistic corrections to the thermal Sunyaev-Zeldovich effect for clusters of galaxies which has a 1\% accuracy for the crossover frequency region where the thermal thermal Sunyaev-Zeldovich effect signal changes from negative to positive sign.  The fitting has been carried out for the ranges $0.020 \leq \theta_{e} \leq 0.035$ and $0 \leq X \leq 15$, where $\theta_{e} \equiv k_{B}T_{e}/m_{e}c^{2}$, $X \equiv \hbar \omega/k_{B}T_{0}$, $T_{e}$ is the electron temperature, $\omega$ is the angular frequency of the photon, and $T_{0}$ is the temperature of the cosmic microwave background radiation.  The overall accuracy of the fitting is better than 0.1\%.  The present analytic fitting formula will be useful for accurate analyses of the thermal Sunyaev-Zeldovich effect for clusters of galaxies.

\end{abstract}


\section{INTRODUCTION}

  In a recent series of papers the present authors (Itoh, Kohyama, \& Nozawa 1998; Nozawa, Itoh, \& Kohyama 1998b; Itoh, Nozawa, \& Kohyama 2000a) have calculated the relativistic corrrections to the thermal, kinematical, and polarization Sunyaev-Zeldovich effects (Zeldovich \& Sunyaev 1969; Sunyaev \& Zeldovich 1972, 1980a, 1980b, 1981).  Other recent references on this subject include Rephaeli (1995); Rephaeli \& Yankovitch (1997), Stebbins (1997), Challinor \& Lasenby (1998,1999), Sazonov \& Sunyaev (1998a, 1998b, 1999).  The present authors have also calculated the relativistic thermal bremsstrahlung Gaunt factor for the intercluster plasma (Nozawa, Itoh, \& Kohyama 1998a; Itoh et al. 2000b).

  The present authors (Nozawa et al. 2000) have presented an accurate analytic fitting formula for the fractional distortion of the photon distribution function caused by the thermal Sunyaev-Zeldovich effect which has been calculated by the direct numerical integration of the collision term of the Boltzmann equation (Itoh, Kohyama, \& Nozawa 1998).  They have obtained 0.1\% accuracy of the analytic fitting formula for most of the parameter region  $0.02 \leq \theta_{e} \leq 0.05$ and $0 \leq X \leq 20$, where $\theta_{e} \equiv k_{B}T_{e}/m_{e}c^{2}$, $X \equiv \hbar \omega/k_{B}T_{0}$, $T_{e}$ is the electron temperature, $\omega$ is the angular frequency of the photon, and $T_{0}$ is the temperature of the cosmic microwave background radiation.  However, the relative error in the crossover frequency region where $X \approx 3.9$ amounted in some cases to more than 5\% owing to the fact that the fractional distortion of the photon distribution function caused by the thermal Sunyaev-Zeldovich effect is such a rapidly varying function of $X$ in this region that the excellent overall fitting is extremely difficult.

  In view of the recent spectacular advances in the accuracy of the observation of the Sunyaev-Zeldovich effect for clusters of galaxies (Birkinshaw 1999), it is worthwhile to present a still more accurate fitting formula for the thermal Sunyaev-Zeldovich effect which has an overall 0.1\% accuracy and has a 1\% accuracy for the crossover frequency region $X \approx 3.9$.  In order to achieve this goal we will restrict the range of the electron temperature to $\theta_{e} \equiv k_{B}T_{e}/m_{e}c^{2} \leq 0.035$.  This temperature range covers all the observed galaxy clusters. 

  The present paper is organized as follows.  In $\S$ 2 we give the the analytic fitting formula.  Concluding remarks will be given in $\S$ 3.

\section{ANALYTIC FITTING FORMULA}

  Itoh, Kohyama, \& Nozawa (1998) have calculated the fractional distortion of the photon distribution function caused by the thermal Sunyaev-Zeldovich effect $\Delta n(X)/n_{0}(X)$ by numerical integration of the collision term of the Boltzmann equation:
\begin{eqnarray}
\frac{\Delta n(X)}{n_{0}(X)} & = & y \, F(\theta_{e}, X) \, ,   \\
X  &  \equiv  &  \frac{\hbar \omega}{k_{B}T_{0}}  \, ,  \\
y & \equiv & \sigma_{T} \, \int d \ell \, N_{e}  \, ,     \\
\theta_{e} & \equiv & \frac{k_{B}T_{e}}{m_{e}c^{2}} \, ,
\end{eqnarray}
where $\omega$ is the angular frequency of the photon, and $T_{0}$ is the temperature of the cosmic microwave background radiation, $\sigma_{T}$ is the Thomson scattering cross section, $N_{e}$ is the electron number density, $T_{e}$ is the electron temperature, and the integral in equation (2.3) is over the path length of the galaxy cluster.  Nozawa et al. (2000) have presented an analytic fitting formula for $F(\theta_{e}, X)$.  Since $F(\theta_{e}, X)$ is a rapidly varying function of $X$, it was extremely difficult to obtain an excellent overall fitting.  Nozawa et al.'s (2000) fitting formula has a general accuracy of 0.1\%.  However, in the crossover frequency region where $X \approx 3.9$, the relative error in some cases amounted to more than 5\%.  In view of the recent spectacular advances in the accuracy of the observation of the Sunyaev-Zeldovich effect for clusters of galaxies (Birkinshaw 1999), it is worthwhile to present a still more accurate fitting formula.

  In the observation of the thermal Sunyaev-Zeldovich effect, the directly observed quantity corresponds to the distortion of the spectral intensity
\begin{eqnarray}
\Delta I & \equiv & \frac{X^{3}}{e^{X} -1} \frac{\Delta n(X)}{n_{0}(X)}  \nonumber  \\
& = & y \, \frac{X^{3}}{e^{X} -1} \,  F(\theta_{e}, X)  \, .
\end{eqnarray}
In this paper we will present an analytic fitting formula for $\Delta I$ instead of $\Delta n(X)/n_{0}(X)$.

  We express the fitting formula for $0.020 \leq \theta_{e} \leq 0.035$, $0 \leq X \leq 15$ as follows:
\begin{eqnarray}
\frac{\Delta I}{y} & = &  \frac{1}{y} \, \frac{X^{3}}{e^{X} -1} \, \frac{\Delta n(X)}{n_{0}(X)}  \,  \nonumber  \\
&  =  & \frac{X^{3}}{e^{X} -1} \, \frac{\theta_{e} X e^{X}}{e^{X}-1} \, \left(
Y_{0} \, + \, \theta_{e} Y_{1} \, + \, \theta_{e}^{2} Y_{2} \, + \,  \theta_{e}^{3} Y_{3} \, + \,  \theta_{e}^{4} Y_{4} \, \right) \, + \,  R    \, .
\end{eqnarray}
The functions $Y_{0}$, $Y_{1}$, $Y_{2}$, $Y_{3}$, and $Y_{4}$ have been obtained by Itoh, Kohyama, and Nozawa (1998) with the Fokker-Planck expansion method, and their explicit expressions have been given.  The graph of the redidual function $R$ is shown in Figures 1, 2.  We take the fitting formula for $R$ as
\begin{eqnarray}
R & = & \left\{ \begin{array}{ll}  0 \, ,  &  \mbox{for $0 \leq X < 2.4$}  \\
        \displaystyle{ \sum_{i,j=0}^{10} \, a_{i \, j} \, \Theta_{e}^{i} \, Z^{j} } \, ,  & 
      \mbox{for $2.4 \leq X \leq 15.0$  \, , }
         \end{array}    \right. 
\end{eqnarray}
where
\begin{eqnarray}
\Theta_{e} & \equiv & \frac{200}{3} \left( \theta_{e} \, - \, 0.02 \right)  \, \, , \, \, \, \, \, \, 0.020 \leq \theta_{e} \leq 0.035  \, ,  \\
Z  & \equiv &  \frac{1}{6.3} \left( X \, - \, 8.7 \right)  \, \, , \, \, \, \, \, \, 2.4 \leq  X  \leq  15.0 \, .
\end{eqnarray}
The coefficients $a_{i \, j}$ are presented in Table 1.  The accuracy of the fitting formula (2.6), (2.7) is generally better than 0.1\%.  The fitting formula has 1\% accuracy in the crossover frequency region where $X \approx 3.9$.  For $\theta_{e} < 0.02$, the results of Itoh, Kohyama, \& Nozawa (1998) give sufficiently accurate results (the accuracy is generally better than 1\% for $0 \leq X \leq 15$).

\section{CONCLUDING REMARKS}

  We have presented a precision analytic fitting formula for the spectral intensity distortion caused by the thermal Sunyaev-Zeldovich effect which has been calculated by the direct numerical integration of the collision term of the Boltzmann equation (Itoh, Kohyama, \& Nozawa 1998).  The spectral intensity distortion corresponds to the directly observed quantity.  The fitting has been carried out for the ranges $0.020 \leq \theta_{e} \leq 0.035$, $0 \leq X \leq 15$.  The accuracy of the fitting is generally better than 0.1\% and is about 1\% in the crossover frequency region where $X \approx 3.9$.  The present results will be useful for accurate analyses of the thermal Sunyaev-Zeldovich effect for clusters of galaxies.  For galaxy clusters with relatively low temperatures $\theta_{e} < 0.02$, the Fokker-Plank expansion results of Itoh, Kohyama, \& Nozawa (1998) will be sufficiently accurate (the accuracy is generally better than 1\%).

\acknowledgements

  We thank Professor Y. Oyanagi for allowing us to use the least-squares fitting program SALS.  This work is financially supported in part by a Grant-in-Aid of Japanese Ministry of Education, Science, Sports, and Culture under contract 10640289.

\newpage


\references{} 
\reference{} Birkinshaw, M. 1999, Physics Reports, 310, 97
\reference{} Challinor, A., \& Lasenby, A., 1998, ApJ, 499, 1
\reference{} Challinor, A., \& Lasenby, A., 1999, ApJ, 510, 930
\reference{} Itoh, N., Kohyama, Y., \& Nozawa, S. 1998, ApJ, 502, 7
\reference{} Itoh, N., Nozawa, S., \& Kohyama, Y. 2000a, ApJ, 533, 588
\reference{} Itoh, N., Sakamoto, T., Kusano, S., Nozawa, S., \& Kohyama, Y. 2000b, ApJS, 128, 125
\reference{} Nozawa, S., Itoh, N., Kawana, Y., \& Kohyama, Y. 2000, ApJ, 536, 31
\reference{} Nozawa, S., Itoh, N., \& Kohyama, Y. 1998a, ApJ, 507, 530
\reference{} Nozawa, S., Itoh, N., \& Kohyama, Y. 1998b, ApJ, 508, 17
\reference{} Rephaeli, Y. 1995, ApJ, 445, 33
\reference{} Rephaeli. Y., \& Yankovitch, D. 1997, ApJ, 481, L55
\reference{} Sazonov, S. Y., \& Sunyaev, R. A. 1998a, ApJ, 508, 1
\reference{} Sazonov, S. Y., \& Sunyaev, R. A. 1998b, Astron. Lett. 24, 553
\reference{} Sazonov, S. Y., \& Sunyaev, R. A. 1999, MNRAS, 310, 765
\reference{} Stebbins, A., 1997, preprint (astro-ph/9705178)
\reference{} Sunyaev, R. A., \& Zeldovich, Ya. B. 1972, Comments Astrophys. Space Sci., 4, 173
\reference{} Sunyaev, R. A., \& Zeldovich, Ya. B. 1980a, ARA\&A, 18, 537
\reference{} Sunyaev, R. A., \& Zeldovich, Ya. B. 1980b, MNRAS, 190, 413
\reference{} Sunyaev, R. A., \& Zeldovich, Ya. B. 1981, Astrophys. Space Phys. Rev., 1, 1
\reference{} Zeldovich, Ya. B., \& Sunyaev, R. A. 1969, Ap\&SS, 4, 301


\newpage
\centerline{\bf \large Figure Captions}

\begin{itemize}

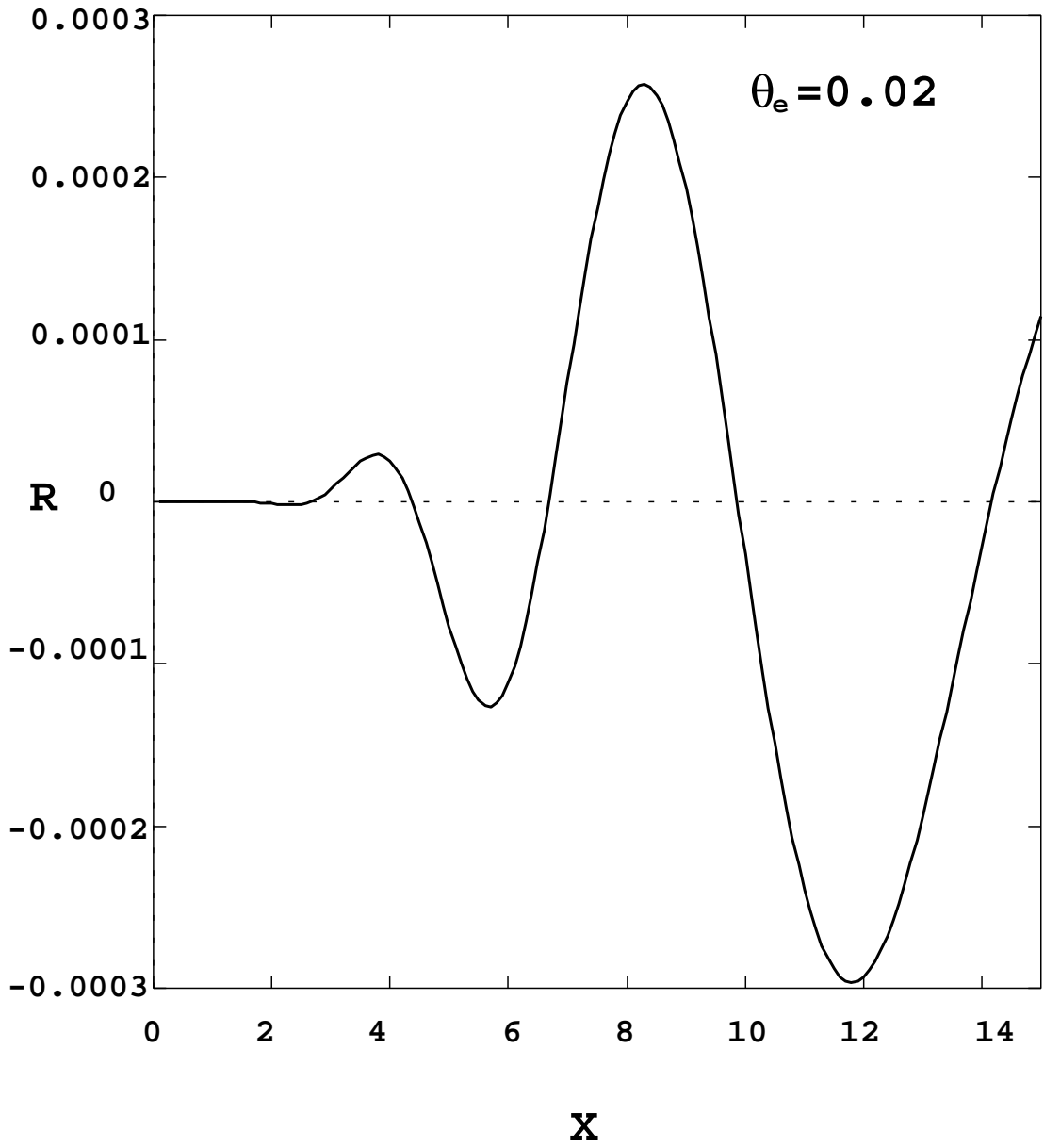
\item Fig.1. The residual function $R$ in equation (2.6) as a function of $X$ for $\theta_{e}=0.02$.

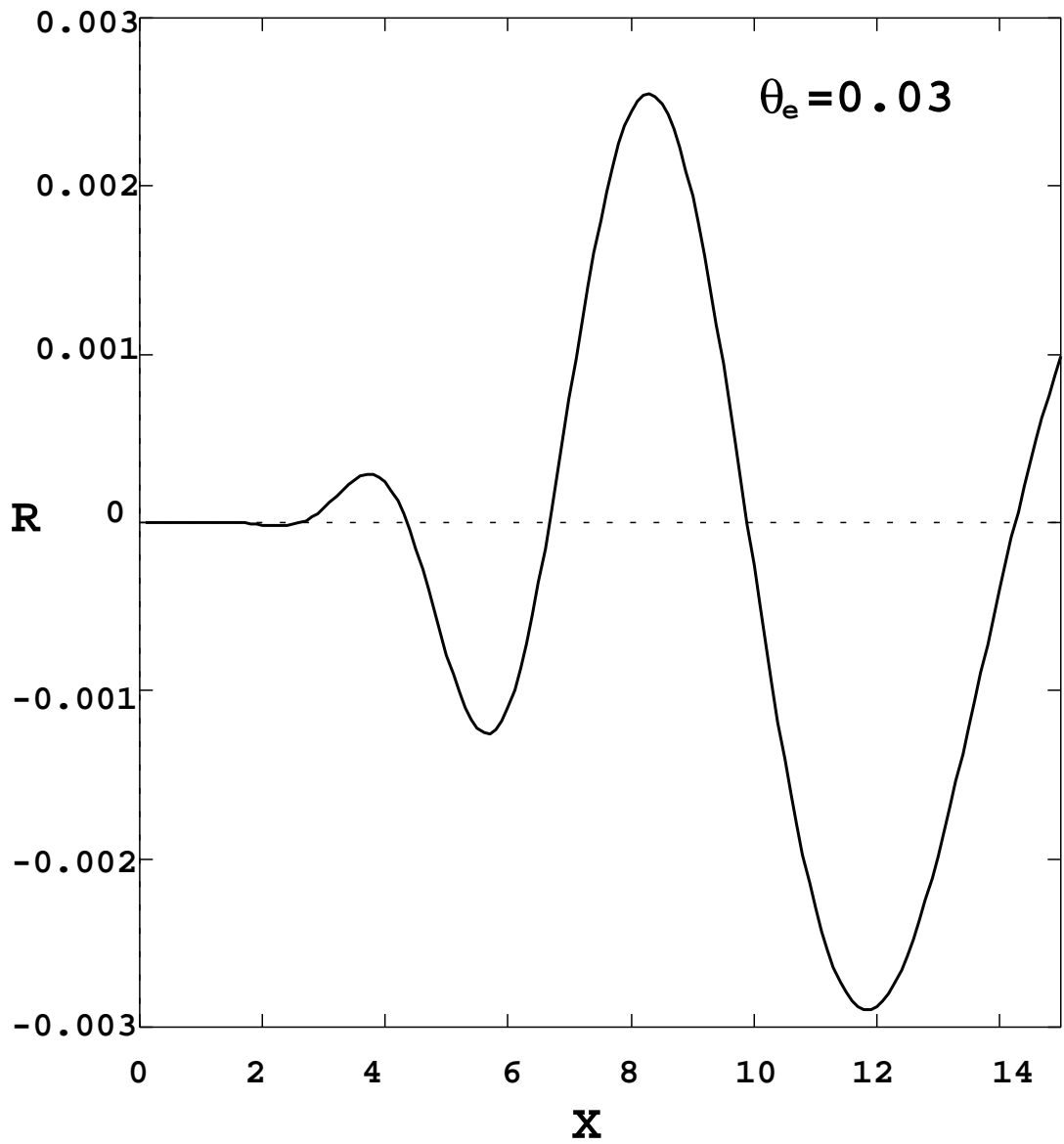
\item Fig.2. Same as Fig.1, but for $\theta_{e}=0.03$.

\end{itemize}

\end{document}